\documentclass{article}

\usepackage[a4paper, total={6in, 9in}]{geometry}

\usepackage[sorting=none]{biblatex}

\usepackage{graphicx}
\usepackage{authblk, subcaption, amssymb}

\DeclareCaptionSubType * [alph]{table}
\providecommand{\makecustomtitle}
{ 
    \date{} 
    \maketitle 
    \vspace{-.25cm} 
}

\providecommand{\keywords}[1]
{ 
    \vspace{.25cm}
    { \small \textbf{\textit{Keywords ---}} #1}
}

\title{ \textbf{Deep vessel segmentation based on a new combination of vesselness filters} }

\author[1]{Guillaume Garret\thanks{guillaume.garret@uca.fr}}
\author[2]{Antoine Vacavant\thanks{antoine.vacavant@uca.fr}}
\author[3]{Carole Frindel\thanks{carole.frindel@creatis.insa-lyon.fr}}
\affil[1,2]{ {\small 
Université Clermont Auvergne,
CNRS, SIGMA Clermont,
F-63000, Clermont-Ferrand, France
}}

\affil[3]{ {\small
Univ Lyon, INSA‐Lyon,
Université Claude Bernard Lyon 1,
UJM-Saint Etienne, CNRS, Inserm,
CREATIS UMR 5220, U1294,
F‐69100, Lyon, France
}}

\begin{document}

\makecustomtitle

\section{Abstract}
Vascular segmentation represents a crucial clinical task, yet its automation remains challenging. Because of the recent strides in deep learning, vesselness filters, which can significantly aid the learning process, have been overlooked. This study introduces an innovative filter fusion method crafted to amplify the effectiveness of vessel segmentation models. Our investigation seeks to establish the merits of a filter-based learning approach through a comparative analysis. Specifically, we contrast the performance of a U-Net model trained on CT images with an identical U-Net configuration trained on vesselness hyper-volumes using matching parameters. Our findings, based on two vascular datasets, highlight improved segmentations, especially for small vessels, when the model's learning is exposed to vessel-enhanced inputs.

\keywords{Segmentation, Deep learning, Vessels, Vesselness filters, Liver, Brain}

\section{Introduction}
\label{sec:intro}
The vascular system, by supplying nutrients and oxygen to the body's organs and removing waste products, plays a crucial role in maintaining the body's metabolic balance \cite{VascularSystem}. Various pathologies are directly linked to the circulatory system and can have severe consequences for organs. Additionally, understanding the distribution of blood vessels is essential for planning surgical procedures. Therefore, the digital reconstruction of the vascular network offers significant benefits for both diagnosis and preoperative planning. Despite its importance, the segmentation process remains primarily manual in practice, and the automation of this labor-intensive task has been a long-standing area of interest.

Efforts to reconstruct vascular structures have traditionally relied on algorithms such as active contour methods \cite{TradiVesselSeg1, TradiVesselSeg2} and model-based approaches like graph-cut techniques \cite{TradiVesselSeg3}. A common preprocessing step involves enhancing vessel contrast using vesselness functions. For instance, in the work by \cite{TradiVesselSeg3}, Frangi's vesselness function is applied prior to a graph-cut, while \cite{TradiVesselSeg2} extends Frangi's function and extracts vessels using a level-set approach based on the Chan-Vese model. 

Traditional rule-based methods have now been complemented by deep learning \cite{nnUNet}. Kitrungrotsakul et al. \cite{VesselNet_Kitrungrotsakul} employ three DenseNet models to extract hepatic vessel features from different planes, namely sagittal, coronal, and axial. Rougé et al. \cite{CascadedTopologicalUNET_Rouge} segment vessels using a topological loss that leverages vascular skeletons. Yan et al. \cite{LVSNet_Yan} introduce LVSNet, a 3D U-Net incorporating multi-scale feature fusion blocks in both encoding and decoding paths, along with spatial attention modules in skip connections replacing conventional concatenation.

In recent research efforts, the scientific community has turned its attention to leveraging the intricate tree-like structures of vascular networks, ushering in innovative methods that integrate Graph Neural Networks into segmentation workflows. This approach, exemplified by \cite{CNN-GAT_Zhang} and \cite{VGN_Shin}, introduces a dynamic interplay between a fully connected network, a graph attention network, and the inherent branching patterns of vessels. Simultaneously, researchers, as highlighted in \cite{UNetVesselness_Affane} and \cite{EffectVesselnessUNet_Survarachakan}, have been investigating the impact of vesselness filters on U-Net-based models. Notably, the latter study delves deeper into filter combinations through composite learning, where vesselness filters are strategically amalgamated, and multi-model learning, employing distinct filter types with subsequent result fusion.

Building on these advances, our work introduces a new paradigm by merging the knowledge of graph-based approaches for evaluation and filter-based approaches for data pre-processing before training. We propose a unique fusion method where vesselness filters are seamlessly integrated into the learning process of a deep neural network. This approach allows our model to discern and exploit the distinctive features enhanced by various filters, such as the Jerman filter's reinforcement of responses at bifurcations \cite{Vesselness_Jerman}. Unlike previous studies limited to specific filter types, especially those based on Hessian matrices, our work pioneers a more comprehensive exploration. By conducting experiments on diverse datasets, including liver (CTA) and brain (MRA), we showcase the versatility of our approach in improving the segmentation of vessels, particularly the finer structures. Furthermore, our evaluation methodology extends beyond conventional evaluation, introducing a novel partitioning approach inspired by \cite{LVSNet_Yan} but adapted to our unique context. Importantly, we introduce a pioneering assessment of topological areas, such as bifurcations, in deep vessel segmentation, providing a more nuanced understanding of model performance.

\section{Materials}
\subsection{Data}
\subsubsection{IRCAD}
The 3Dircadb-01 dataset, abbreviated as IRCAD, comprises 20 publicly available contrast-enhanced CT thoracic scans. The dataset maintains an equal gender distribution, with 75\% of patients exhibiting liver tumors. 

The original dimensions of the coronal and sagittal planes remain constant at 512 voxels, while the axial plane dimension varies from 74 to 260. Voxel sizes are not uniform across patients and axes, ranging from 0.56 to 0.87 mm in the $\vec{x}$ and $\vec{y}$ directions, and from 1.0 to 4.0 mm in the $\vec{z}$ direction.

In the work of~\cite{VesselnessBenchmark}, images are resampled to homogenize voxel spacing. Each data undergoes resampling to its optimal resolution using a third-order B-spline interpolation. For instance, an image with voxel spacing (0.65 mm, 0.65 mm, 1.5 mm) is resampled to (0.65 mm, 0.65 mm, 0.65 mm).

We opt for this variant of IRCAD to ensure consistency with the application of vesselness filters (see Section \ref{subsec:vesselness_filter})

\subsubsection{Bullitt}
The original Bullitt dataset \cite{bullitt} comprises a collection of 100 (T1, MRA) image pairs from healthy volunteers. However, only a portion of the dataset has been annotated for vessel segmentation purposes. For our experiments, we have access to 33 MRA images with sparsely annotated regions that have been manually refined \cite{VesselnessBenchmark}. The voxel spacing for the MRA images is (0.51 $\times$ 0.51 $\times$ 0.80 mm), with a volume size of (448 $\times$ 448 $\times$ 128 voxels).

Similar to our approach with the IRCAD dataset, we utilize the resampled version of the Bullitt dataset, aligning with its finest resolution through third-order B-spline interpolation \cite{VesselnessBenchmark}.

\subsubsection{Vesselness filters} \label{subsec:vesselness_filter}
In a significant effort, \cite{VesselnessBenchmark} introduces a benchmark evaluating key vesselness filters, including Frangi, Jerman, Sato, Zhang, Meijering, and RORPO (Ranking the Orientation Responses of Path Operators). The first five filters gauge tubularity based on the Hessian matrix, while RORPO adopts a morphological approach.

\begin{figure}[!t]
\centering
\includegraphics[scale=0.25]{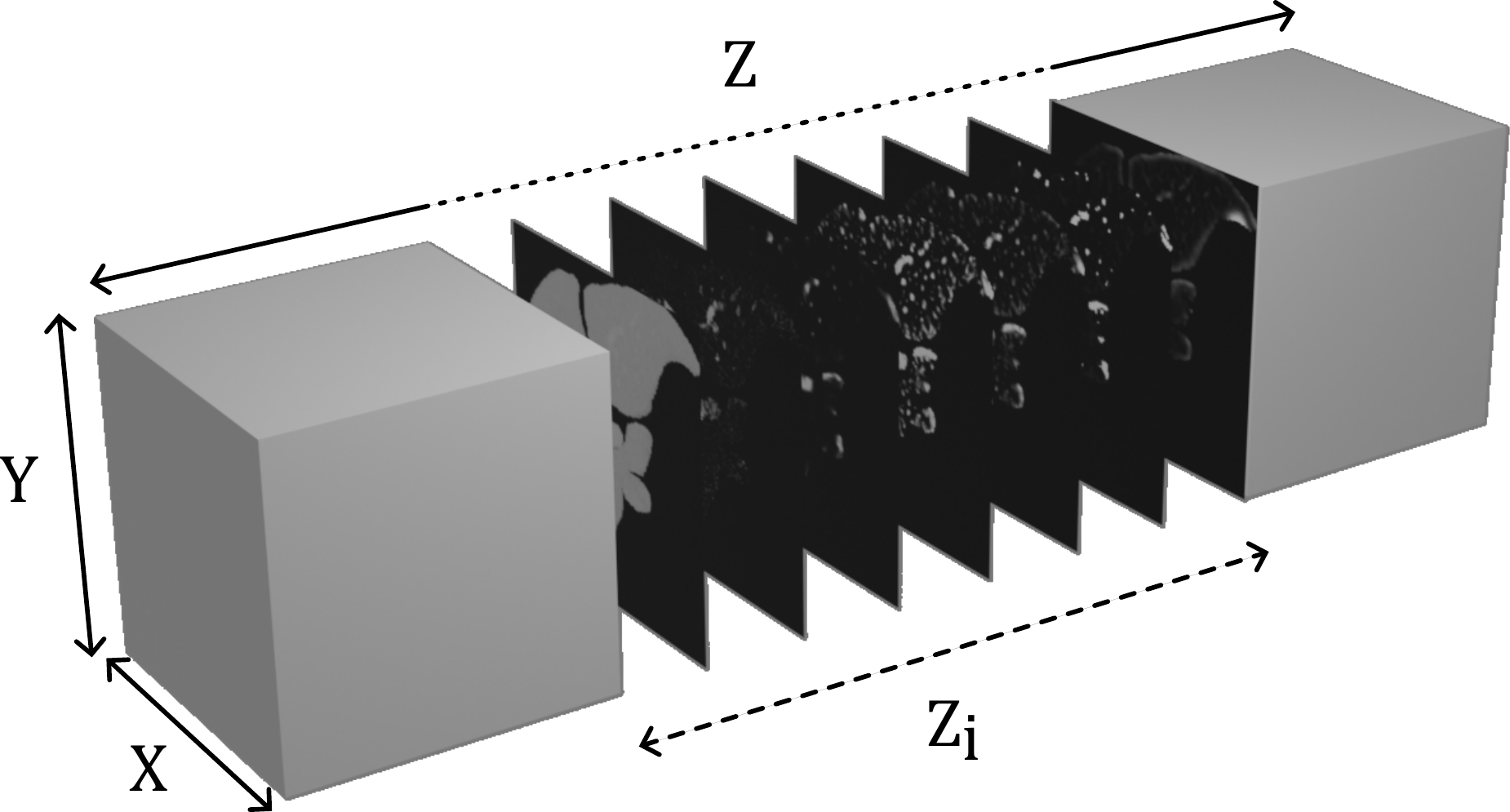}
\caption{Hyper-volume where filters are concatenated to the original scan, with each $Z_{i}$ depth corresponding to a different filter (Original, Frangi, Jerman, Sato, Zhang, Meijering, RORPO, respectively).}
\label{fig:hypervolume}
\end{figure}

Building on the insights from this benchmark, we generate a vessel-enhanced version for the two datasets mentioned earlier. Here, the filters are concatenated to their respective original scans in the channel dimension, as illustrated in Fig.~\ref{fig:hypervolume}.

\subsection{Experimental setup}

The U-Net model, sourced from the MONAI implementation, comprises 4 depth levels and trained on a workstation equipped with a Quadro RTX 8000 GPU. The number of feature maps increases by a factor of 2 at each level, ranging from 16 to 128 at the deepest level. Training employs the Adam optimizer with an initial learning rate of $10^{-4}$. The loss function integrates a weighted combination of Dice loss and Focal loss, with equal weights assigned to each term. Patches of size (64 $\times$ 64 $\times$ 64) are extracted from the input images, with a 25\% overlapping ratio. To address class imbalance, we ensure that at least 85\% of the patches contain a foreground voxel. Data augmentation techniques, including random rotation, random flip, and Gaussian smoothing, are also applied.

To validate the generalizability of the U-Net, we adopt a cross-validation approach. For the Bullitt dataset, 15\% of the data is allocated for the test phase, while the remaining 85\% is divided into 5-folds for training and validation. In contrast, due to the smaller size of IRCAD, a dedicated test set is considered impractical. Instead, the entire dataset is split into 5-folds, with one sample from the validation subset dedicated as a test volume for each fold model. Despite its unconventional nature, we believe this hybrid approach, combining k-fold and leave-one-out cross-validation, offers a reasonable compromise considering the constraints of the IRCAD dataset. To enhance the robustness of our results, we also conduct an evaluation of the U-Net on the validation sets for IRCAD.

\section{Experiments}
\subsection{Training}
To assess the influence of vesselness filters on segmentation, we conducted experiments on two distinct datasets: IRCAD and Bullitt. For each dataset, we executed two sets of experiments—one using the original data and another with the vessel-enhanced data (e.g., IRCAD, IRCAD-enhanced, Bullitt, and Bullitt-enhanced). This approach facilitated a comprehensive examination of the impact of vesselness filters on segmentation across both datasets. Our hyperparameters closely align with those employed in two prior studies \cite{EffectVesselnessUNet_Survarachakan}.

\subsection{Evaluation}
To assess the effectiveness of our models, we initially employ the Dice score. While the Dice score evaluates the overlap between a segmentation map and a reference, it tends to be more sensitive to larger structures than to thinner ones, presenting challenges given the morphology of vessels.

To address this sensitivity imbalance, we refine our evaluation by considering vessel size partitions, following the methodology outlined in \cite{VesselnessBenchmark}. The ground-truth undergoes a skeletonization process through a distance transform, generating a graph representing the vascular tree. Each identified branch is labeled uniquely, and a subsequent labeling is performed, assigning the maximum size of each branch as its label. Subsequently, all branches are grouped into classes—small, medium, and large—based on specific vessel size intervals, as detailed in Table~\ref{tab:vessel_size_paritions}. Notably, in the case of Bullitt's vessels, only two groups—small and medium—are used, owing to the lower variability in vessel size.

The final masks, denoted as $M_{large}$, $M_{medium}$, and $M_{small}$, are derived through a dilation process, employing a ball structuring element whose diameter adjusts based on the vessel size. This dilation method ensures the continuity between vascular masks.

\begin{table}[!ht]
\begin{center}
\caption{Vessel partitions according to vessel size in $mm$}
\label{tab:vessel_size_paritions}
\begin{tabular}{|r || c | c | c|}
\hline & Small & Medium & Large \\
\hline IRCAD & $[0,3]$ & $]3,6]$ & $]6,+\infty[$ \\
\hline Bullitt & $[0,0.513]$ & $]0.513, +\infty[$ & $\varnothing$ \\
\hline
\end{tabular}
\end{center}
\end{table}

We further evaluate segmentations at vessel bifurcations. Here, bifurcations are localized given the adjacency matrix of the skeleton, then, the $M_{bif}$ mask is obtained by dilating the bifurcation voxels and intersecting the resulting image with the ground-truth, following the method described in \cite{VesselnessBenchmark}.

Additionally, in the context of the vessel segmentation task, preserving the topology of structures is crucial, a factor not considered by the Dice score. Addressing this concern, we introduce the clDice score \cite{clDice_Shit}. Unlike the traditional Dice score, the clDice metric utilizes the vessel skeleton to assess the preservation of vessel connectivity.

\section{Results}

\subsection{IRCAD}
\begin{figure}[!htb]
    \centering
    
    \includegraphics[scale=1.5]{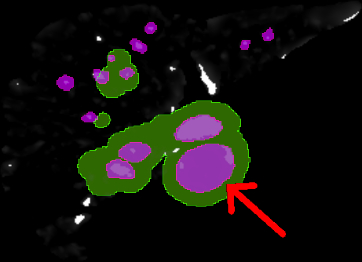}
    \caption{Signal loss for a large vessel. White: enhanced vessels; purple:ground truth; green:mask $M_{large}$.}
    \label{fig:largeVessel_signalLoss}
\end{figure}

Our analysis on the IRCAD dataset, detailed in Table~\ref{subtab:res_ircad}, reveals a notable improvement in the mean Dice for small vessels with the application of enhancement filters, with an average deviation over +9\%. However, a negative impact on large vessels is observed, likely due to signal loss in our enhanced dataset (see Fig.~\ref{fig:largeVessel_signalLoss}). Adjusting the scale space of filters may mitigate this.

Bifurcation segmentation slightly decreased in the test set, traced back to a specific sample's drop in Dice score under our vesselness configuration.

Additionally, vessel-enhancement improved the preservation of vascular topology, reflected in increased clDice scores across partitions. On test sets, improvements ranged from an average deviation of +1.47\% (medium vessels) to +16\% (small vessels), and on validation sets, from +4.16\% to +19.7\%. Visual representation is provided in Fig.~\ref{fig:ircad_bullit_qual}.

\begin{table}[!htb]
    \label{tab:full_results}
    \caption{Mean and standard deviation for Dice and clDice scores, according to vessel partitions.}

    \begin{subtable}{\linewidth}
    
        \centering
        \caption{IRCAD} \label{subtab:res_ircad}
    
        \begin{tabular}{|l||ll|ll|}
            \hline Dice  & \multicolumn{2}{c|}{Original} & \multicolumn{2}{c|}{Vessel enhanced} \\ \hline
             & \multicolumn{1}{c}{Test} & \multicolumn{1}{c|}{Validation} & \multicolumn{1}{c}{Test} & \multicolumn{1}{c|}{Validation} \\ \hline
            Global & $\mathbf{0.631\pm0.044}$ & $\mathbf{0.645\pm0.021}$ & $\mathbf{0.638\pm0.071}$ & $\mathbf{0.672\pm0.021}$ \\ 
            Large & $\mathbf{0.745\pm0.094}$ & $\mathbf{0.775\pm0.015}$ & $\mathbf{0.708\pm0.082}$ & $\mathbf{0.667\pm0.018}$ \\ 
            Medium & $\mathbf{0.645\pm0.086}$ & $\mathbf{0.629\pm0.059}$ & $\mathbf{0.668\pm0.092}$ & $\mathbf{0.695\pm0.032}$ \\ 
            Small & $\mathbf{0.430\pm0.054}$ & $\mathbf{0.389\pm0.053}$ & $\mathbf{0.522\pm0.04}$  & $\mathbf{0.515\pm0.029}$ \\ 
            Bifurcations & $\mathbf{0.728\pm0.139}$ & $\mathbf{0.809\pm0.036}$ & $\mathbf{0.718\pm0.124}$ & $\mathbf{0.853\pm0.041}$ \\ \hline
        \end{tabular}
        
        \vspace{.2cm}

        \begin{tabular}{|l||ll|ll|}
            \hline clDice & \multicolumn{2}{c|}{Original} & \multicolumn{2}{c|}{Vessel enhanced}\\ \hline
             & \multicolumn{1}{c}{Test} & \multicolumn{1}{c|}{Validation} & \multicolumn{1}{c}{Test} & \multicolumn{1}{c|}{Validation}\\ \hline
            Global & $\mathbf{0.614\pm0.039}$ & $\mathbf{0.623\pm0.061}$ & $\mathbf{0.720\pm0.104}$ & $\mathbf{0.763\pm0.027}$ \\ 
            Large & $\mathbf{0.828\pm0.059}$ & $\mathbf{0.822\pm0.028}$ & $\mathbf{0.860\pm0.049}$ & $\mathbf{0.864\pm0.025}$ \\ 
            Medium & $\mathbf{0.733\pm0.086}$ & $\mathbf{0.751\pm0.077}$ & $\mathbf{0.748\pm0.165}$ & $\mathbf{0.852\pm0.029}$ \\ 
            Small & $\mathbf{0.483\pm0.071}$ & $\mathbf{0.458\pm0.087}$ & $\mathbf{0.645\pm0.051}$ & $\mathbf{0.655\pm0.037}$ \\ 
            Bifurcations & $\mathbf{0.797\pm0.067}$ & $\mathbf{0.805\pm0.061}$ & $\mathbf{0.866\pm0.054}$ & $\mathbf{0.899\pm0.014}$ \\ \hline
        \end{tabular}
    \end{subtable}

    \vspace{.5cm}
    
    \begin{subtable}{\linewidth}        
        \centering
        \caption{Bullitt} \label{subtab:res_bullitt}

        \begin{subtable}{.5\linewidth}
            \begin{tabular}{|l||l|l|}
                \hline Dice & \multicolumn{1}{c|}{Original} & \multicolumn{1}{c|}{Vessel enhanced}\\ \hline
                Global & $0.642\pm0.004$ & $\mathbf{0.656\pm0.002}$  \\ 
                Medium & $0.709\pm0.012$ & $\mathbf{0.748\pm0.008}$  \\ 
                Small & $0.685\pm0.005$ & $\mathbf{0.699\pm0.003}$  \\ 
                Bifurcations & $0.888\pm0.005$ & $\mathbf{0.899\pm0.004}$ \\ \hline
            \end{tabular}
        \end{subtable}

        \vspace{.2cm}
    
        \begin{subtable}{.5\linewidth}
            \begin{tabular}{|l||l|l|}
                \hline clDice & \multicolumn{1}{c|}{Original} & \multicolumn{1}{c|}{Vessel enhanced}\\ \hline
                Global & $0.775\pm0.005$ & $\mathbf{0.783\pm0.002}$  \\ 
                Medium & $0.760\pm0.014$ & $\mathbf{0.791\pm0.008}$  \\ 
                Small & $0.831\pm0.006$ & $\mathbf{0.844\pm0.004}$  \\ 
                Bifurcations & $0.935\pm0.003$ & $\mathbf{0.942\pm0.003}$ \\ \hline
            \end{tabular}
        \end{subtable} 
    \end{subtable}
\end{table}

\subsection{Bullitt}
\begin{figure}
\centering
    \includegraphics[width=.25\textwidth]{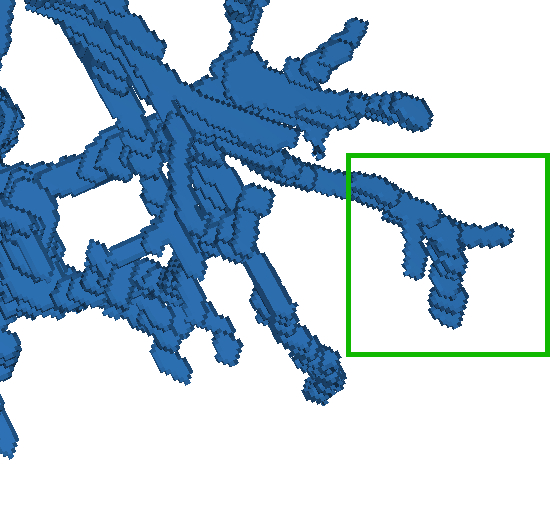}
    \includegraphics[width=.25\textwidth]{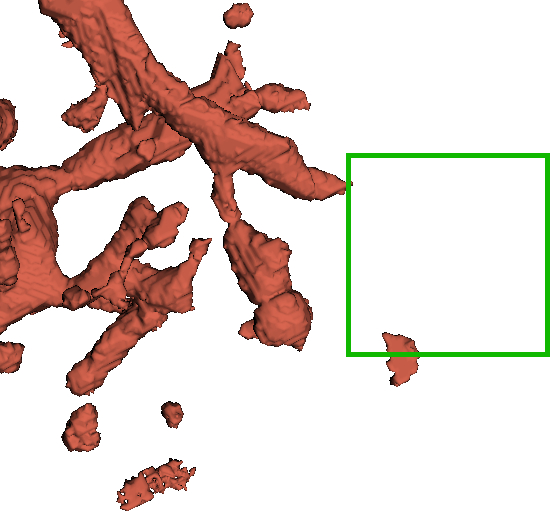}
    \includegraphics[width=.25\textwidth]{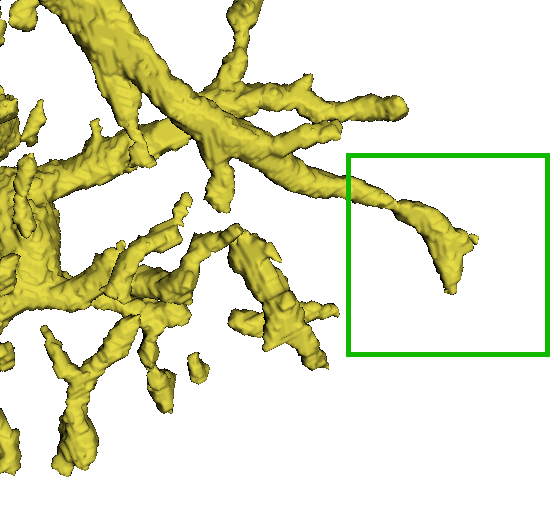}

    \includegraphics[width=.25\textwidth]{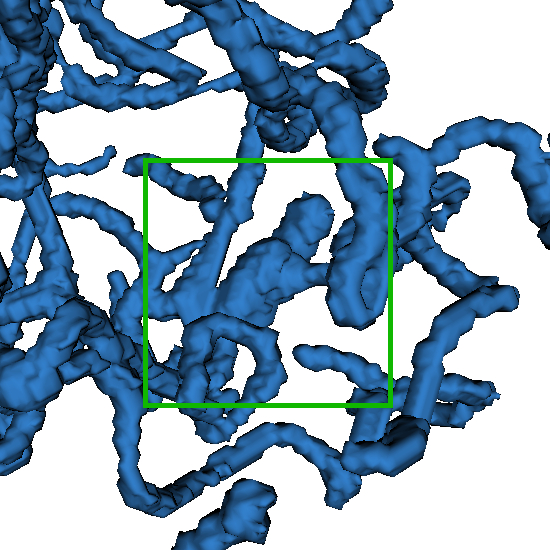}
    \includegraphics[width=.25\textwidth]{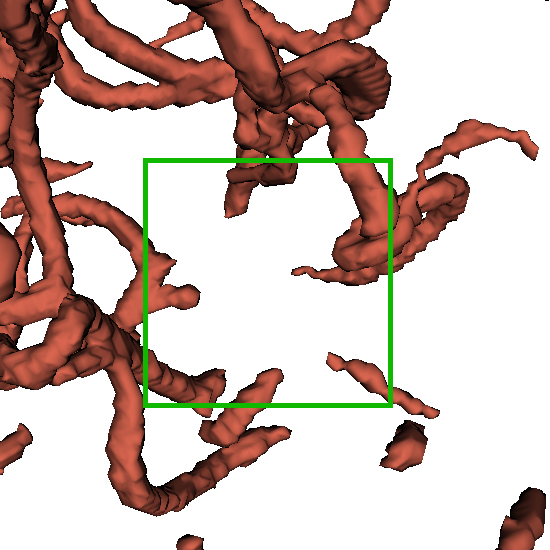}
    \includegraphics[width=.25\textwidth]{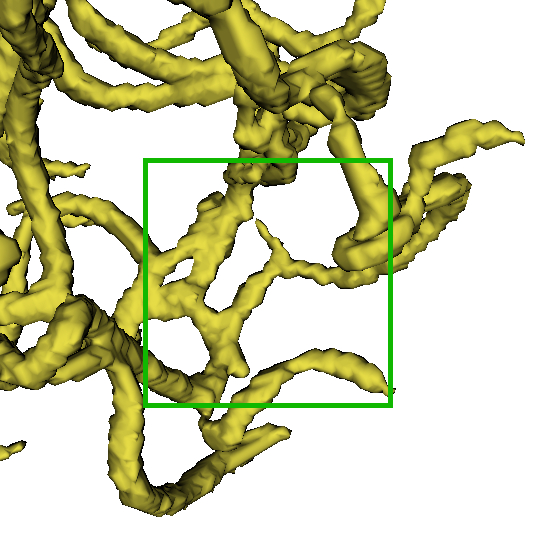}

    \caption{Example of vessel segmentation for IRCAD (left) and Bullitt (right) datasets. Top to bottom: Ground-truth, U-Net's output using original images, and U-Net's output using our proposed hyper-volumes.}

\label{fig:ircad_bullit_qual}
\label{fig:ircad_enhanced}
\end{figure}

For the Bullitt dataset, we have improved segmentation with regard to our both metrics, as shown in Table~\ref{subtab:res_bullitt}. This is consistent with the previous results, as the Bullitt's vessels are significantly smaller than those of IRCAD. Hence we are not impacted by large vessels signal loss. In addition, we note that the vesselness benefit is less for Bullitt, from $+1.09\%$ to $+3.88\%$, for bifurcations and medium vessels respectively. This smaller average deviations can be explained by the fact that the Bullitt dataset is almost noise-free compared to the IRCAD, which reduces the impact of the filters. Indeed, the Peak Signal-to-Noise Ratio (PSNR) for the original Bullitt of $19.04\pm0.43$ improved to just $21.56\pm1.12$ after enhancement, compared to IRCAD, which increase significantly from $9.35\pm1,22$ to $19.74\pm1.60$ after enhancement. An illustration of segmentation is given Figure~\ref{fig:ircad_bullit_qual}.

\section{Conclusion}
In this study, we introduced a novel fusion method for vesselness filters in filter-based learning, concatenating filters into a hyper-volume. Our comparative study, utilizing a U-Net model with fixed parameters, assessed the model's efficiency when trained on original images versus our proposed hyper-volumes. The results yielded valuable insights into the efficacy of our approach.

Notably, the incorporation of vesselness-based learning demonstrated an improvement in the Dice score, particularly for small vessels, and preserved vascular structure topology, as indicated by the clDice measure. We evaluated our approach on two diverse datasets—Bullitt for cerebral MR and IRCAD for hepatic CT applications—following a partitioning approach.

Looking ahead, our future work will delve into examining the contribution of the studied filters using eXplainable AI methods. We believe this approach could reveal the most influential filters for the models, enhancing our understanding of the model's behavior.

\newpage

\section{Acknowledgments}
\label{sec:acknowledgments}

This project receives support from the Région Auvergne-Rhône-Alpes under the DAISIES project. 
We thank the NVIDIA Academic Hardware Grant Program for providing the GPU resources that have been instrumental in carrying out the experiments for this work.
The MR brain images utilized in this study, obtained from healthy volunteers, were collected and provided by the CASILab at The University of North Carolina at Chapel Hill. These images are distributed through the MIDAS Data Server at Kitware, Inc.

\section{Compliance with Ethical Standards}
This research study was conducted retrospectively using human subject data made available in open access by the IRCAD Institute and TubeTK project. Ethical approval was not required as confirmed by the license attached with the open access data.

\printbibliography

\end{document}